\documentclass[12pt]{article}
\usepackage{graphicx}

\newcommand\pubnumber{CIPANP2015-SEKuhn}
\newcommand\pubdate{\today}

\def\odu{Department of Physics\\
Old Dominion University, Norfolk VA 23529 (USA)}
\def\support{\footnote{Work supported by the U.S. Department of Energy, 
          under contract DEFG0296ER40960.}}

\def\Title#1{\begin{center} {\Large #1 } \end{center}}
\def\Author#1{\begin{center}{ \sc #1} \end{center}}
\def\Address#1{\begin{center}{ \it #1} \end{center}}

\newcommand\pubblock{\rightline{\begin{tabular}{l} \pubnumber\\
         \pubdate  \end{tabular}}}
\newenvironment{Abstract}{\begin{quotation}  }{\end{quotation}}
\newenvironment{Presented}{\begin{quotation} \begin{center} 
             PRESENTED AT\end{center}\bigskip 
      \begin{center}\begin{large}}{\end{large}\end{center} \end{quotation}}
\def\Acknowledgements{\bigskip  \bigskip \begin{center} \begin{large}
             \bf ACKNOWLEDGEMENTS \end{large}\end{center}}

\def\beq{\begin{equation}}
\def\eeq#1{\label{#1}\end{equation}}
\def\eeqn{\end{equation}}

\def\beqa{\begin{eqnarray}}
\def\eeqa#1{\label{#1}\end{eqnarray}}
\def\eeqan{\end{eqnarray}}

\let\bar=\overbar

\def\Dslash{\not{\hbox{\kern-4pt $D$}}}
\def\dslash{\not{\hbox{\kern-2pt $\del$}}}

\def\msb{{\bar{\ssstyle M \kern -1pt S}}}

\begin{document}
\begin{titlepage}
\pubblock

\vfill
\Title{Modifications of Nucleons in Nuclei}
\vfill
\Author{ Sebastian E. Kuhn\support}
\Address{\odu}
\vfill
\begin{Abstract}
I summarize recent results and discuss upcoming and planned experiments that attempt to elucidate how the structure of nucleons might be modified by nuclear binding.
\end{Abstract}
\vfill
\begin{Presented}
CIPANP2015\\
Vail, CO,  May 18--23, 2015
\end{Presented}
\vfill
\end{titlepage}
\def\thefootnote{\fnsymbol{footnote}}
\setcounter{footnote}{0}

\section{Introduction}

Nuclear Physics, from its beginnings in the 1930's up to the present day, has had tremendous 
successes describing the masses, energy levels, shapes and other properties of nuclei as well as
their structures and reactions in terms of
microscopic (or collective) models where nuclei are composed of individual nucleons interacting
with each other through the exchange of mesons, perhaps augmented by a hard repulsive core. 
In particular, state-of-the-art
calculations including sophisticated nucleon-nucleon 
potentials~\cite{Machleidt:2000ge,Epelbaum:2008ga,Piarulli:2014bda,Epelbaum:2014sza,Entem:2015xwa}, 
or based on fully relativistic
nuclear models~\cite{Gross:2010qm}, can describe many features of the nuclear response in elastic,
quasi-elastic and inelastic electro-weak scattering off nuclei. While some care has to be taken to incorporate the
required conservation of energy and momentum (as well as current conservation) in the description of
such processes (because of nuclear binding), these models are remarkably successful in spite of ignoring
any internal structure of the nucleons making up nuclei, or at least any modification of that structure
from that of free nucleons.

In stark contrast, we know that the ultimate theory describing  nucleons and nuclei is Quantum
Chromo Dynamics (QCD). In this framework, both free nucleons and nuclei are bound (stationary) systems
containing quarks, antiquarks and gluons interacting through their color couplings. From this perspective, the 
main distinction between a nucleon and a nucleus is just the net difference between the number of quarks and antiquarks, which is equal to $3 A$ ($A$ being the baryon number of a nucleon or nucleus). At first blush,
it is then quite surprising that nuclei should appear as a collection of nucleons at all, let alone (largely)
unmodified ones. However, due to the unique properties of QCD, namely confinement at large distance scales and spontaneous chiral
symmetry breaking, such a ``clustering'' into nucleon-like sub-structures together with meson exchange currents
can be understood at least qualitatively. Yet, it would be very surprising, within this framework, if there were
{\em no} differences between the internal structure of those ``nucleon-like constitutents'' of nuclei and
free nucleons.

A comparison with the (much more precisely calculable) theory of Quantum Electrodynamics (QED) may
illuminate this point. In this theory, the structure and the mass of the electron are ultimately due in part to
its virtual fluctuations into electrons plus photons, additional electrons and positrons, which are suppressed
by powers of the electromagnetic coupling strength, $\alpha_{\mathrm EM} \approx 1/137$. As a consequence,
one should expect, at least in principle, a modification of the structure of an electron bound in, say, a hydrogen
atom vs. that of a free one. Indeed, one can interpret observations such as the Lamb shift in hydrogen as
a consequence of this internal structure modification. However, because the binding energy of the electron
in hydrogen
is only a small fraction of its mass ($\alpha_{\mathrm EM}^2/2 = 0.0027\%$), the effect is tiny
(about a million times smaller than the applicable binding energy) and can
only be observed with the extreme precision available to modern atomic spectroscopy. At the other end of
the scale, there is no question that atoms themselves are significantly modified when bound into molecules -
the eigenstates of the electrons in a hydrogen molecule have completely different shapes than those in
hydrogen atoms. This corresponds to a much larger relative binding energy in this case - H$_2$ is
bound by about 1/3 of the binding energy of free hydrogen atoms.

It seems clear then that for the case considered here, nucleons bound in nuclei, one should expect binding 
effects intermediate between the two extreme examples above, since the binding energy of individual nucleons
in nuclei lies somewhere between 0.12\% and 0.9\% of their masses. 
Indeed, there are several experimental results that appear to unambiguously confirm some modification of 
nucleon structure in bound nucleons; some of these results are discussed in the next section. 
Nevertheless, a quantitative comparison between such experimental signals and 
rigorous QCD-based predictions
remains elusive, due both to the difficulties of reliable QCD calculations in the low-energy regime and due
to the considerably larger uncertainties in hadronic experiments and their interpretation
(including the above-mentioned ``trivial'' binding effects due to energy-momentum and current
conservation), compared to, e.g., atomic spectroscopy. This is the reason for the oft-cited adage
``Nuclear modifications of nucleon structure are like the Mafia in Sicily: everyone knows they are
there, but it is very difficult to find hard evidence''.

\section{Experimental Evidence}

\begin{figure}[htb!]
\centering
\includegraphics[width=5in]{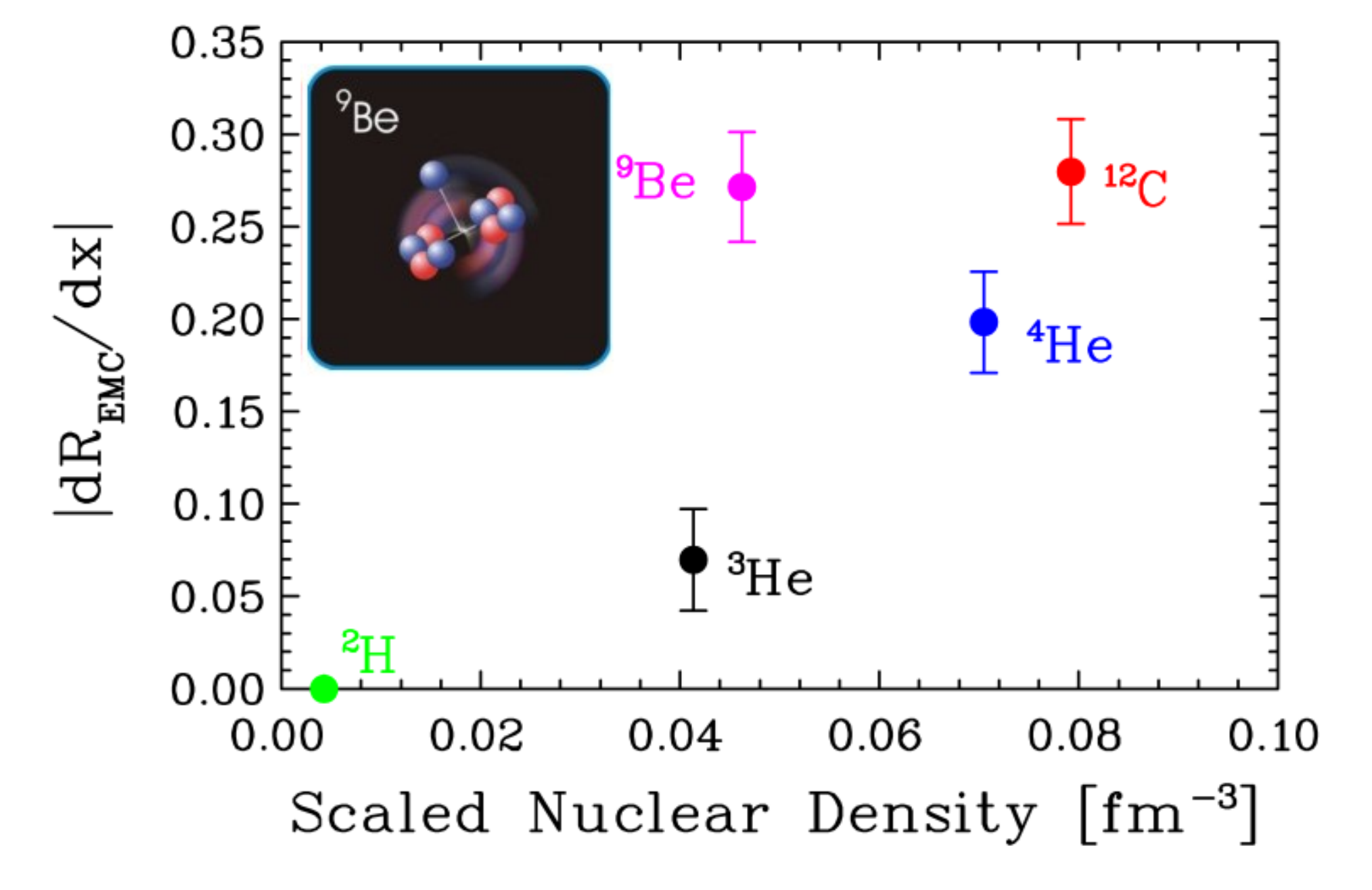}
\caption{\baselineskip 13pt Slope of the ratio between bound and free nucleon structure functions
with the scaling variable $x$, versus the average density of the nuclei investigated in~\cite{Seely}.
Note that here, as usual, the EMC ratio $R_{EMC}$ of structure functions
 is actually defined relative to the deuteron, which is only an approximation to the average of a free
 proton and a free neutron.}
\label{fig:seely}
\end{figure}

One of the earliest experiments showing incontrovertible evidence for the modification of
nucleon structure inside the nucleus is the famous EMC experiment~\cite{EMC} which
has been corroborated by a large number of experiments at many labs~\cite{Dasu, Gomez, NMC, BCDMS}.
These experiments show a significant reduction of the per-nucleon structure function $F_2(x)$
in the valence region, $0.2 \le x \le 0.8$ relative to that of a free nucleon, with a roughly
linear behavior that is universal in shape for all nuclei, but with a slope that tends to increase
for heavier and denser nuclei. A recent precision experiment at Jefferson Lab~\cite{Seely}
has extended our knowledge of this EMC effect towards a range of lighter nuclei, including 
$^{3,4}$He, $^9$Be and $^{12}$C. Surprisingly, it was found that the ``EMC slope'' does not
follow a simple correlation with the average nuclear density of the studied nuclear species.
Instead, as shown in Fig.~\ref{fig:seely}, $^9$Be shows a strong EMC effect somewhere between
$^4$He and $^{12}$C, in spite of being a rather dilute nucleus (see inset in Fig.~\ref{fig:seely}).

A closer inspection shows, however, that 8 out of the 9 nucleons in $^9$Be are located inside tightly 
bound ``alpha clusters'', with only the remaining neutron further out. Hence, the {\em local} 
nuclear density seen by most nucleons in $^9$Be is much higher than the average, global density. This
can be seen as strong evidence that the strength of the EMC effect (the magnitude of the slope)
is determined by the local nuclear environment of the struck nucleon.

\begin{figure}[htb!]
\centering
\includegraphics[trim=0mm 35mm 0mm 40mm, clip=true, width=5in]{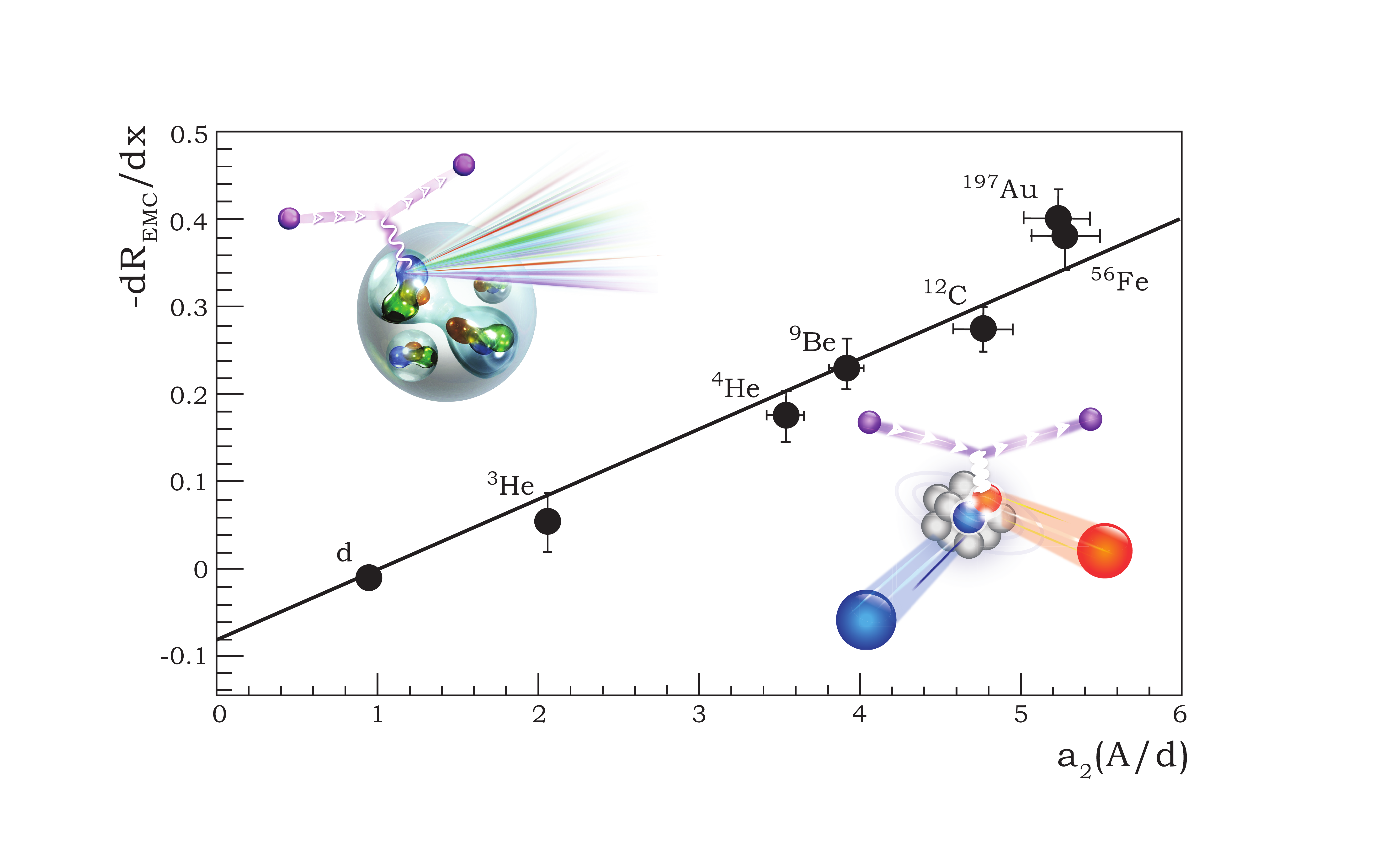}
\caption{ \baselineskip 13pt  
Correlation between the strength (slope) of the EMC effect ($y$-axis) and the relative
probability to find a nucleon inside a high-momentum nucleon correlation ($a_2$) relative to the same
probability in deuterium. $a_2$ measures the relative yield of DIS on a nucleus at large $x > 1.4$,
where only high momentum nucleons can contribute to the strength of the response. A rich body of
experimental evidence shows that these high momentum nucleons are nearly always paired with
a partner of the opposite type and of nearly equal and opposite momentum; see elsewhere in these proceedings.}
\label{fig:Or}
\end{figure}

This conclusion is further corroborated by a comparison with another effect that should directly depend on the local nuclear density, namely the strength of short-range (high-momentum) nucleon-nucleon correlations inside nuclei. Indeed, it was found~\cite{Larry,Or} that this strength correlates nearly perfectly with the strength of the EMC effect over
a wide range of nuclei, as shown in Fig.~\ref{fig:Or}. This remarkable agreement seems to indicate that
either nucleons in short-range (high-momentum) correlations contribute the bulk of the overall EMC
effect (see, e.g., the model by Frankfurt and Strikman~\cite{FS}), or that both the correlation probability and the EMC effect are governed by a common underlying
feature, e.g., the local nuclear density (consistent with ``quark-meson coupling'' models like the one
 in~\cite{TT}). 

\begin{figure}[b!]
\centering
\includegraphics[width=4in]{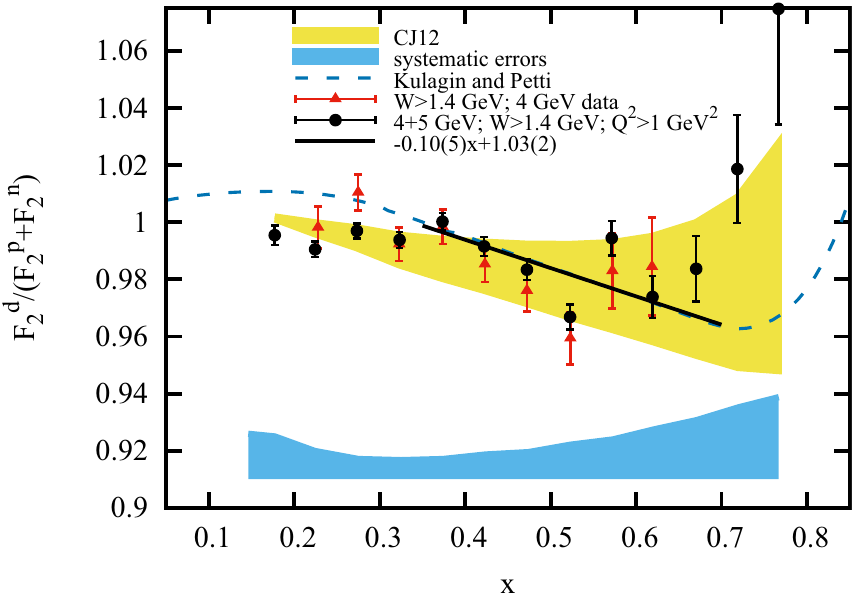}
\caption{\baselineskip 13pt EMC effect in Deuterium~\cite{Keith}. See text for explanation.}
\label{fig:Keith}
\end{figure}

Most recently, a first attempt to measure directly the EMC effect in deuterium has been published~\cite{Keith},
based on the data of the BONuS experiment~\cite{Slava} which, for the first time, attempted to directly
extract the structure function $F_{2n}$ of the free neutron. Within the (rather large) experimental uncertainties,
strongly suggestive evidence of a  EMC slope of order -0.1 was found (see Fig.~\ref{fig:Keith}), which is
consistent with the extension of the straight line fit in Fig.~\ref{fig:Or} to $a_2=0$, corresponding to
free nucleons.

Beyond measurements of DIS on nuclei, experimental hints for nucleon modification in nuclei also come
from form factor measurements on bound nuclei. For instance, a recent measurement~\cite{Paolone} of
the  ratio of
electric to magnetic form factors of a proton bound in $^4$He, relative to a free proton, found a substantial
reduction of this quantity that is not easily explained without invoking modifications of the nucleon
size inside a nucleus. 

\section{Ongoing and Planned Experiments}

Given the fundamental importance of understanding the QCD structure of bound nucleons, a large
array of new experiments are either underway or planned to further study in detail various
aspects of the EMC effect and hopefully clarify the underlying mechanism. These experiments
use a variety of probes to elucidate the dependence of the EMC effect on quark flavor and nucleon type
(proton vs. neutron), on the virtual photon polarization (longitudinal vs. transverse) and on nucleon spin. Further experiments aim to tackle directly the question
whether high-momentum nucleons in the nucleus are more strongly modified than average, ``mean
field'' nucleons. High precision measurements of the free neutron structure will allow us to
directly extract the EMC effect in deuterium, the lightest and least dense nucleus. Finally, further
measurements of in-medium nucleon form factors are ongoing, for instance at the MAMI electron
scattering facility.
 Below we give a brief summary of some of these experiments; many of them are described in
 detail elsewhere in these proceedings.

Among the alternative probes used to study the EMC effect, DIS with neutrinos and Drell-Yan experiments
play a prominent role. Experiments like Miner$\nu$a that study the interaction of neutrino beams with 
nuclear targets open a different window on nucleon modifications in medium, since they are sensitive to different combinations
of quark flavors. In fact, the famous ``NuTeV anomaly''~\cite{Zeller}, while originally interpreted as 
a violation of Standard Model expectations, might indeed be (at least partially) explained~\cite{Cloet} 
by non-trivial isospin-dependent nucleon structure modifications in the iron target used by NuTeV.
Similarly,
Drell-Yan experiments like SeaQuest are particularly well suited to access the largely
unexplored contribution of sea quarks and antiquarks to the EMC effect.

The bulk of new experiments to further study the EMC effect are planned for the energy-upgraded
Jefferson Lab (Thomas Jefferson National Accelerator Facility, Newport News, VA).
The upgrade, which is nearly complete, together with extensive new experimental facilities in
four halls, provides for significantly higher energy (11-12 GeV), high intensity electron and photon
beams well-suited for high-precision experiments in the valence region. These facilities will
be used to continue the study of the ``classical'' EMC effect in DIS on nuclei, with
unprecedented precision, kinematic reach and a plethora of nuclei. In particular,
 experiment E12-10-008 will study, for the first time, many isotopes of the same
elements over a large range in nucleon number A, e.g. $^{3-4}$He, $^{6-7}$Li, $^{10-11}$B and
$^{40,48}$Ca. These data will allow us to unambiguously extract the difference between proton and
neutron structure modifications in the nucleus, in particular once the new free neutron data
from experiments like ``BONuS12'' and ``Marathon'' are available. The latter experiment will also
directly measure the EMC effect for the isospin pair $^3$He/$^3$H. These results will be augmented
with precision measurements of the ratio $R$ for longitudinal vs. transverse virtual photon absorption
on nuclei.

\begin{figure}[htb!]
\begin{minipage}{0.5\linewidth}
\includegraphics[width=\linewidth]{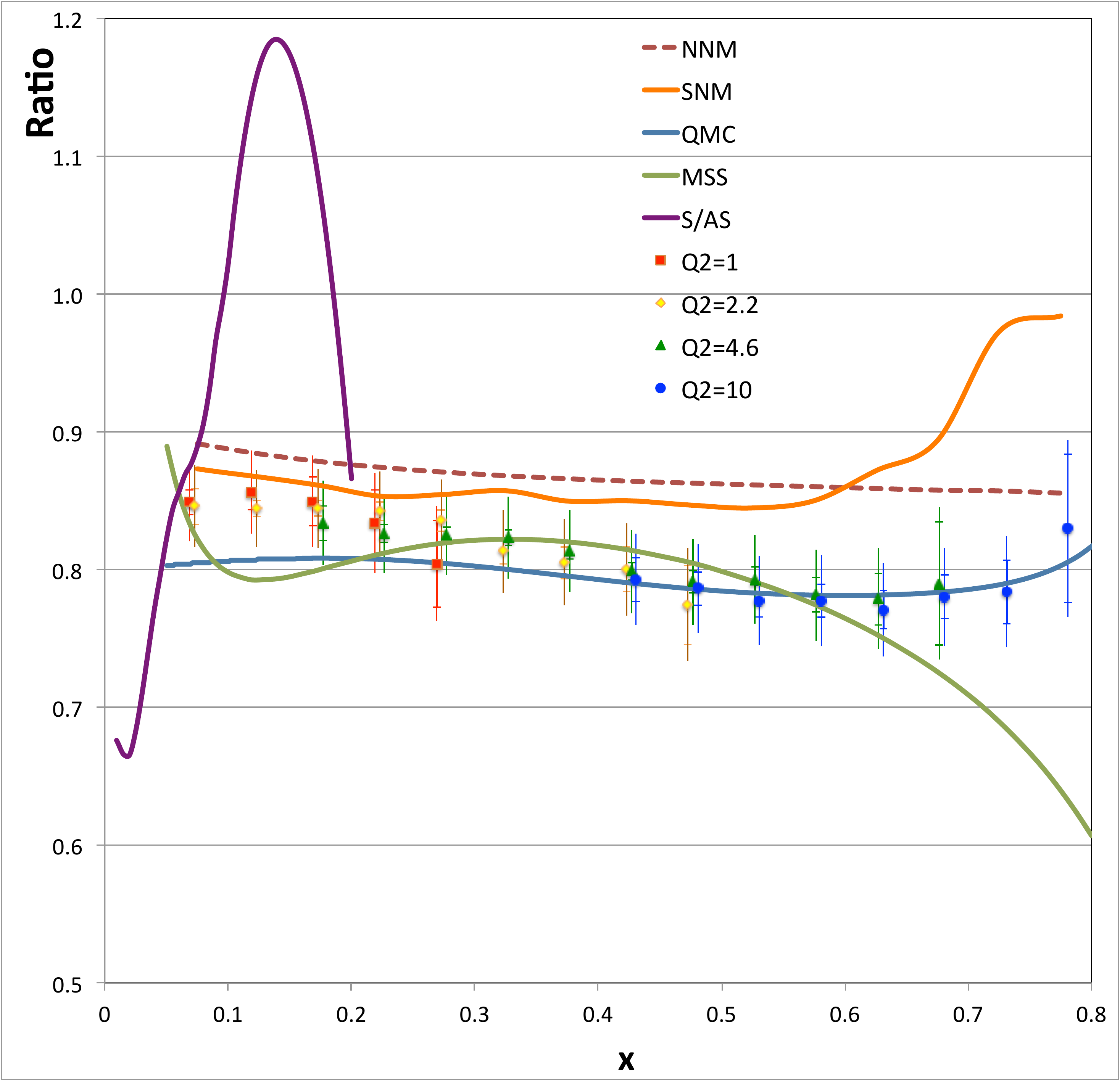}
\end{minipage}
\hfill\begin{minipage}{0.5\linewidth}
\includegraphics[width=\linewidth]{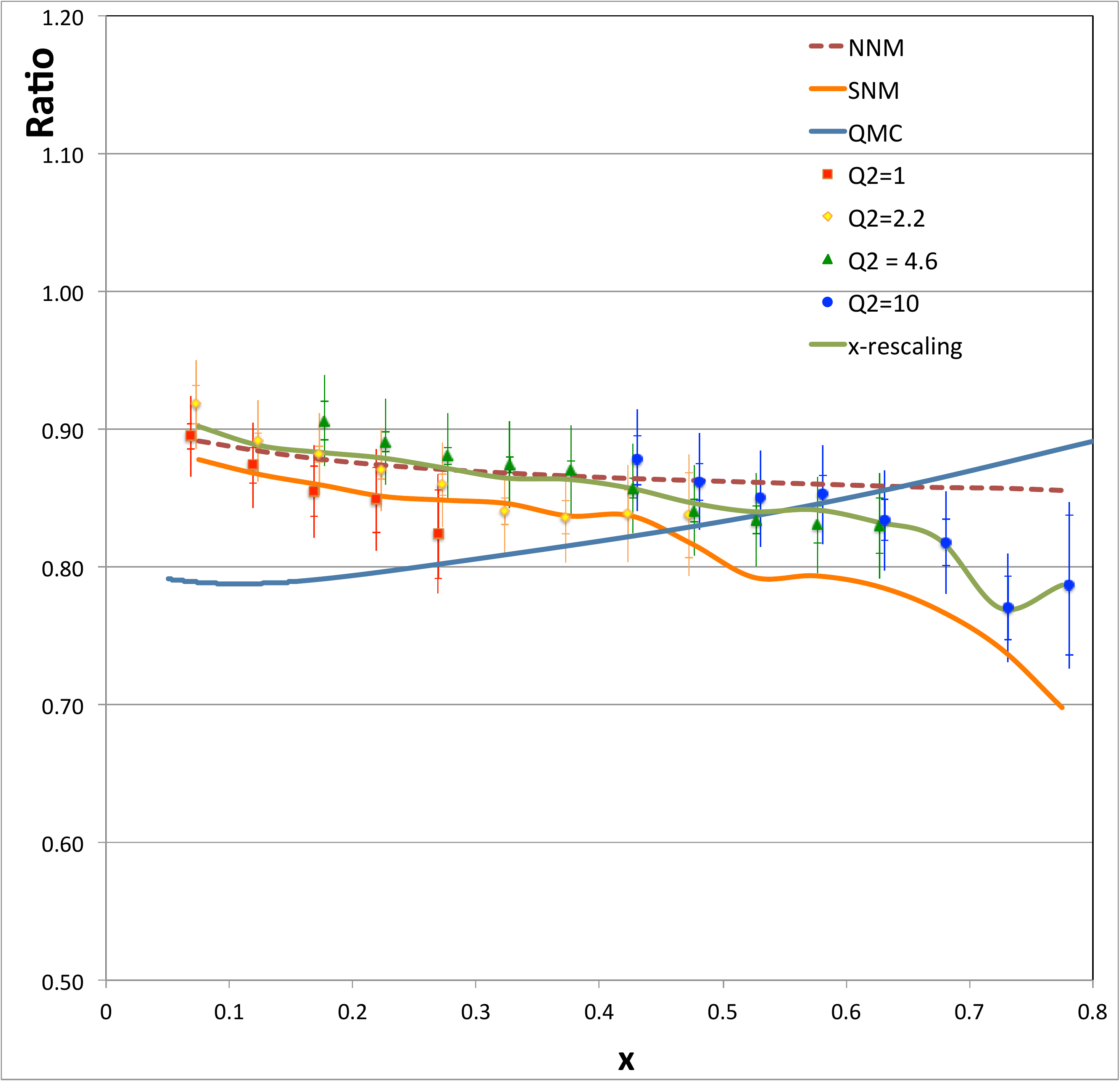}
\end{minipage}
\caption{\baselineskip 13pt Expected results for experiment E12-14-001 at Jefferson Lab.
The l.h.s. shows the ratio of spin structure functions $g_1$ for $^7$Li over the free proton, while
the r.h.s. shows the same ratio for the asymmetries $A_1$. The inner error bars representred
the expected statistical uncertainties, and the outer ones 
include the systematic uncertainties.
The various model curves shown are from a simple additive ``nucleons-only'' model without EMC effect and
with (SNM) and without (NNM) accounting for nuclear Fermi motion, as well as a ``Quark-Meson Coupling''
type model~\cite{Cloet2} (QMC) and two alternative models~\cite{MSS,SAS}.}
\label{fig:Seb}
\end{figure}

Another degree of freedom that has not been experimentally explored so far at all is the spin
of the nucleon. While the EMC was also the first (of many) experiments to find
 a surprisingly small contribution of quark spins to the
overall nucleon spin, so far we don't have any information on the ``squared EMC effect'', namely
the modification of nucleon spin structure functions for nucleons bound in nuclei. Such
measurements could play a decisive role distinguishing between various models of the EMC
effect, since they make strikingly different predictions for the ratio of bound to free 
spin structure functions. Experiment E12-14-001 will measure the ratio of the spin structure function
$g_1(x)$ and the asymmetry $A_1(x)$ for the nucleus $^7$Li over the free proton. Figure~\ref{fig:Seb}
shows the expected results compared to various model expectations.

Further studies involving polarized nuclei aim to measure the tensor-polarization structure function $b_1$.
There are also plans to study Deeply Virtual Compton Scattering from nuclei (in particular $^4$He) to find
evidence for modifications of Generalized Parton Distributions  (GPDs), following a first proof-of-principle experiment
with Jefferson Lab's 6 GeV beam. Since GPDs encode both the longitudinal momentum and transverse
spatial distribution of quarks, they can be used to test models where the EMC effect is due to nucleon
``swelling'' in the medium.

Finally, several experiments will use the technique of ``spectator tagging'' to directly study
DIS on fast-moving nucleons inside the nucleus. This technique was pioneered with 6 GeV
experiments like ``BONuS''~\cite{Slava} and involves the simultaneous detection of a scattered
electron and a backwards-going spectator nucleon. BONuS applied this method, selecting 
slow-moving protons from a deuterium target, to tag nearly on-shell neutrons and measure their
free structure functions. On the other hand, by measuring high momentum spectators, one can
ensure that the electron scattering took place on a nucleon that was part of a short-range correlation,
thereby accessing any possible enhancement of the EMC effect in such nucleon pairs.
This will be exploited by experiment E12-11-107 (with proton spectators) and a companion
experiment looking for fast backward neutrons, both with deuterium targets. Beyond that,
plans are underway to extend this technique to heavier nuclei.

\section{Outlook and Conclusions}
Given the mounting experimental evidence, there can be hardly any disagreement that non-trivial 
nuclear binding effects on nucleon structure exist. Upcoming experiments, in particular at the
energy-upgraded Jefferson Lab, will further sharpen this evidence and map out their detailed properties.
A complete description of these nuclear effects is a necessary part of our understanding of the
microscopic (QCD) structure of nucleons {\em and} nuclei, and therefore the manifestation of QCD in all
strongly bound systems. It is also an important input for the interpretation of experiments that
rely on nuclear targets to study fundamental physics - from neutrino scattering (see the 
famous NuTeV anomaly that is now considered to be at least partially due to nuclear effects) to
the measurement of structure functions of the neutron, which by necessity involves nuclear targets.
We can look forward to a new era of precision measurements, with commensurate
advances in theoretical understanding, in this area. In particular, recent progress in 
ab-initio calculations of QCD bound-state properties on the lattice make fully microscopic
models of nucleons {\em and} nuclei appear feasible in the not-too-distant future.

Ultimately, an electron-ion-collider as proposed by the US nuclear physics community will
be required to complete our picture of parton distributions in momentum and space, both
in nucleons and in nuclei. In particular, such a machine will be uniquely suited to study
the modification of gluon and sea quark distributions in a wide range of nuclei, and
further elucidate the phenomenon of shadowing. 

While we haven't quite solved yet the
30-year old puzzle posed by the original EMC measurement, we can finally look forward to
a definite answer provided by the new accelerators, experimental methods and
theoretical advances already in place or on the horizon.

\Acknowledgements
My research is supported by the U.S. Department of Energy under grant
DE-FG02-96ER40960.

\end{document}